# Method of indirect estimation of default probability dynamics for industry-target segments according to the data of Bank of Russia

Mikhail Pomazanov*
Ph.D. Assistant Professor
Higher School of Economics National Research University, Russia
*Presenting author: mhubble@yandex.ru

**Abstract**

A direct method for calculating default rates by industry and target corporate segments is not possible given the lack of statistical data. The proposed paper considers a model for filtering the dynamics of the probability of default of corporate companies and other borrowers based on indirect data on the dynamics of overdue debt supplied by the Bank of Russia. The model is based on the equation of the balance of total and overdue debts, the missing links of the corresponding time series are built using the Hodrick–Prescott filtering method. In retail lending segments (mortgage, consumer lending), default statistics are available and supplied by Credit Bureaus. The presented method is validated on this statistic. Over a historical limited period, validation has shown that the result is trustworthy. The resulting default probability series are exogenous variables for macro-economic modelling of sectoral credit risks.

## 1. Introduction

Numerous global financial crises have shown that the reason for the significant increase in arrears on loans is an increase in the probability of default (Probability of Default, PD) of borrowers, and the resulting losses banks have to cover at their own expense. In order to minimize losses from the growing credit risk, any credit organizations need effective ways of quantifying the probability of default, as well as the dynamics of this probability. Stress testing requires macroeconomic models to explain this dynamic. An external benchmark is required to estimate this probability $PD_i$ at any given time i.

The first option and the most obvious is the benchmark based on CB[1] statistics, optimal for the portfolio of individuals. As a reliable benchmark, it is wise to choose the default rate supplied by CB, such as NCB[2]. However, CB statistics almost significantly do not cover the corporate segment including the average business. In this segment, there are no full-fledged default statistics. And, even if the absolute value of the number of defaults can somehow be obtained from legal sources, the value of the denominator, which takes into account only active enterprises, credited by banks and leading economic activities, is difficult to determine. In addition, legal decisions are very late regarding the event of default. CB in Russia has historically not been directed to service the credit history of legal entities, there were no laws obliging a legal entity to keep a credit history in CB.

---

[1] CB – Credit Bureaus
[2] The National Bureau of Credit Stories delivers a quarterly printed (and electronic) edition of the National Credit Bulletin to its subscribers for credit report (credit organizations).

In practice, developed countries, where there are also problems in full-fledged statistical sources, analysts monitor the frequency of default on the reports of rating agencies of the Big Three[3], but studies on Russia are not conducted separately, too small rating coverage of the pool of corporate enterprises and groups of companies. The situation with Russian rating agencies is improving, for example, for several years the Expert RA[4] has been publishing a matrix of defaults[5] and keeps no zero statistics of defaults since 2008, the reliability of which is increasing. However, at the moment the statistical error of the benchmark $PD_i$ on the data Expert RA is still significant, but soon this benchmark may soon become objective.

There are sources of data for the benchmark $PD_i$ that are not direct, but only indirect, but quite objective. These are data on the levels of reservation and the level of delinquency[6] in the banking system as a whole and in terms of industries. In article (Kuznecov K.B., et all, 2011)) several approaches were proposed for transforming these data into PD. However, there are significant drawbacks to these approaches, such as the lack of structure data, such as reserves. Namely, it is not clear what level of realized losses is hidden in the structure of reserves. This is important because PD is a probability, so you are interested in the reserve level except for the hopeless ones. The second important point is the restoration of reserves and overdue due to reverse processes of repayment of loans, exit from default, write-off, etc. This requires a meaningful model that is sensitive to dynamics. This model is proposed in this work.

There are still market methods for assessing the market probability of default based on bond, equity and CDS prices. Many of them are already classic, for example, based on the Merton model (Merton, R.C., 1974), (McSwown, J.A., 1993). KMV (currently part of Moody's Analytics) has developed hybrid approaches (Sobehart, J. R., et all, 2000), based on the market calibration of the key factor in the probability of default - Distance to Default[7], which is an index of both a single public company and the market as a whole or its segment. A broad overview of practical and theoretical approaches is presented in the paper (Lapshin V. A., Smirnov S. N., 2012), which proposes a method of information in one assessment obtained by different ways of assessments of default probability, risk-neutral and real. Two "engineering" ways of translating risk-neutral probabilities into real ones through a communication equation derived from certain

---

[3] See, for example, "Annual Default Study: Corporate Default and Recovery Rates", Moodyes Report-Annual Period. Or "Default, Transition, and Recovery: Annual Global Corporate Default And Rating Transition Study", StandardandPoors Report. Period - Annual.

[4] The first certified national rating agency of Russia

[5] Historical data on default levels on the rating categories of the rating scales used as of the date, the source of the Expert RA, the periodic of the report - six months.

[6] Source Bank of Russia,, «Outstanding amount (including overdue debt) of loans granted to resident legal entities and individual entrepreneurs, by economic activity and use of funds (as of reporting date)», https://cbr.ru/statistics/bank_sector/sors/

[7] Distance to Default is an indicator of the distance to default, associated with the probability that the market value of a firm's assets will fall below the value of its debt. In order to realize the face value of the debt, an equal amount of short-term liabilities is accepted plus half of the long-term liabilities derived from the balance sheet data. The model is then calibrated using the market value of the firm and the observed volatility in the price of its shares.

considerations are considered.

A direct method of calculating the frequency of defaults on industry and target corporate segments is not possible in the absence of statistics. The proposed work considers a model of filtering the dynamics of the probability of default of corporate companies and other borrowers based on indirect data on the dynamics of arrears supplied by the Bank of Russia. The model is based on the balance equation of aggregate and overdue debt, the missing connections of the respective time series are built by the Hodrick-Prescott filtration method (Hodrick R. Prescott E. C., 1997) (commonly known as HP filter). In retail lending segments (mortgage, consumer lending) default statistics are available and supplied by CB. This statistic validated the method presented. At the historical limited interval, validation showed that the result was credible. The resulting series of probability of default are exogenous variables for macro-economic modeling of industry credit risks.

## 2. Filtering method

The first step is to establish a balance sheet that simulates a change in the level of arrears. This equation is as follows:

$$NPL_{i+1} - NPL_i = P_i \cdot (E_i - NPL_i) - R_i \cdot NPL_i \tag{1}$$

$i = 1 \dots N$ , historic interval month number

$E_i$ – debt in the industry segment

$NPL_i$ – overdue debt in the industry segment

$P_i$ – indicator of default rate per month i, $1 \geq P_i > 0$

$R_i$ – Monthly recovery share indicator

The next step is to introduce functionality that filters dependencies $P_i$ and $R_i$, the basic filtering requirements are pretty obvious:

Continuity $P_i$

The $R_i$ convergence to the average

Filtering functionality is being built, which is analogous to the HP filter:

$$\sum_{i=2}^{N-1} \left( \ln\left(\frac{1}{P_{i+1}} - 1\right) + \ln\left(\frac{1}{P_{i-1}} - 1\right) - 2 \cdot \ln\left(\frac{1}{P_i} - 1\right) \right)^2 + \lambda \cdot \sum_{i=1}^{N} (R_i - RR)^2 \to min_{\{P_i, R_i, i=1\dots N\}} \tag{2}$$

The model allows you to build a solution: $\hat{P}_i(RR, \lambda)$, which depends on two unknown parameters of the $RR, \lambda$.

The average annual probability of default, which is a model of the frequency of realized defaults (DF) is based on the Bayes formula

$$PD_i = 1 - \prod_{k=i-11}^{i} (1 - \hat{P}_k(RR, \lambda))$$

## Conditions for determining unknown parameters:

$\sum_{i=n}^{N} PD_i(RR, \lambda) = N \cdot PD\_TTC$, где $PD\_TTC = $ DF

*at the interval $i = n \dots N$, $n > 11$ and is defined by the economic cycle;*

The number of m lows or highs of $M^k$, such as

$$PD_{M^{k-1}}(RR, \lambda) \geq PD_{M^k}(RR, \lambda) \leq PD_{M^{k+1}}(RR, \lambda) \ or$$

$$PD_{M^{k-1}}(RR, \lambda) \leq PD_{M^k}(RR, \lambda) \geq PD_{M^{k+1}}(RR, \lambda),$$

$k = 1 \dots m$ corresponds to the number of minimums or maximums observed directly or indirectly by a number of $DF_i$, among the possible values $\lambda$, the maximum is chosen.

The $RR, \lambda$ parameters are calculated once for the overall segment (e.g. "all industries" segment) and are unchanged (i.e. constants) for sub-segments (industries).

<u>Example:</u> Statistics on loans provided to legal entities - residents and individual entrepreneurs in RUR, by types of economic activity and individual areas of use of funds are used.

<u>Initial data:</u> Number of minimums m=3 (period: Oct. 2010-October 2019) PD TTC[8] (October 2010-October 2019 according to Expert RA) is 3.49%.

Result:  RR = 34,42% , $\lambda$ = 0,015625

## 3. Validation of the model

Validation of the model should be carried out on the segment for which there is objective data on the frequency of defaults. The NCB statistical bulletin on the one hand and the data on the delay on the other are taken as supporting data. It is necessary to ensure the similarity of credit market segments. NCB data provides a time series of overdue terms of 90 days or above in retail lending, mortgage, car loans and credit card loans.

Statistics of the Bank of Russia provides information on loans provided to individuals - residents, as well as debt (including arrears) on housing loans granted to individuals- residents. If you subtract the last two rows of one of the other, you get an analogue of conventional retail loans except mortgages. NCB data on auto loans, credit cards and retail should be summarized, then this series of $PD_i$ is compared with the data of the Bank of Russia, subject to calibration on the average PD TTC and the choice of the option of $\lambda$ corresponding to an equal number of minimums (maximums) for the period of existence of open suppressing data (since 2012).

The result of validation is that the model (1), (2) gives the ranks of quasi-PD (indirect PD) that are close to the real DF (with a determination rate of $R^2 = 95 - 99\%$)). Therefore, there is reason to trust this model.

## 4. Industry calculations

Model (1), (2) is applied on the data of the source of the Bank of Russia "Debt, including overdue, on loans granted to legal entities - residents and individual entrepreneurs, by types of economic activity and individual areas of use of funds" provided the average $PD_i$ average value according to the Expert RA. It is clear from the calculations that the dynamics of the probability of default is very different for different industries both in terms of the amplitude of fluctuations, and in terms of the emerging problems with overdue in the industry. From the data for 2009-2010, you can see the depth associated with the global crisis, there is a sense associated with the events of 2014-2015.

---

[8] TTC is through the cycle a standard designation of the average long-term PD for a period no less than an economic cycle.

The calculations presented in Table 1 show a significant stratification of risks in industries.

*Table 1 . The results of calculating the average PD by industry for the period April 2009 - October 2019, as well as the coefficient of variation equal to the ratio of the Standard Deviation to the Average Value.*

| Industry | The average | Variation Ratio (Standard Deviation to Average Value) |
|---|---|---|
| Total | 3.6% | 0.3 |
| mining | 2.4% | 1,2 |
| - extraction of fuel and energy minerals | 2.9% | 1.4 |
| manufacturing industries | 3.7% | 0,4 |
| -production of food products, including beverages, and tobacco | 5.1% | 0.3 |
| - wood processing and production of wood products | 14.0% | 0.5 |
| -pulp and paper production; publishing and printing activities | 6.7% | 0.7 |
| -production of coke, oil products and nuclear materials | 5.7% | 1.1 |
| -chemical production | 2.2% | 0.7 |
| -production of other non-metallic mineral products | 7.4% | 0.8 |
| -metallurgical production and production of finished metal products | 3.6% | 0.6 |
| - manufacture of machinery and equipment | 3.9% | 0.5 |
| - manufacture of machinery and equipment for agriculture and forestry | 9.3% | 0.8 |
| -production of vehicles and equipment | 2.6% | 1.1 |
| -car production | 8.0% | 1,2 |
| production and distribution of electricity, gas and water | 2.0% | 0.9 |
| agriculture, hunting and forestry | 4.5% | 0.2 |
| - agriculture, hunting and provision of services in these areas | 4.5% | 0.2 |
| building | 7.4% | 0.6 |
| - construction of buildings and structures | 8.1% | 0.6 |
| transport and communication | 2.5% | 0.7 |
| air transport activities, obeying and not obeying the schedule | 10.5% | 1,2 |
| wholesale and retail trade; repair of vehicles, motorcycles, household goods and personal items | 4.8% | 0,4 |
| real estate transactions, rental and service provision | 4.2% | 0.5 |
| other activities | 3.1% | 0.5 |
| to complete settlements | 2.9% | 0.5 |

It is necessary to make only an important reservation that the traditional for the Bank of Russia break-up by types of economic activity is not significantly uniform, so some industries (e.g., wood processing and production of wood products) are not comparable in terms of the volume of lending and the number of enterprises (for example, the wholesale and retail trade industry has the maximum volume of activity and number of enterprises, but is not divided into sub-sectors).

### 5. On building a time series of default probabilities and preparing a macro model

Each bank has its own market niche of credit business, expressed in industry specifics. Therefore, in order to build a series of PD equivalent to the market, it is necessary to draw up the ranks of $E_i$ , $NPL_i$ required for the application of the PD filtering model weighted by industry shares corresponding to the bank's portfolio. The second step is to establish the average DF for a long period according to the relevant statistics of the Bank's portfolio. The third step uses a filtering model and gives the appropriate dynamics $PD_i$ . On the fourth step is the macro model of this series.

Table 2 presents the result of building a regression macro model, corresponding to the industry specifics of the bank's sub-portfolio.

*Table 2. The characteristics of macro models built on the ranks of PD, filtered by the model (1), (2) on the data of the Bank of Russia, taking into account the industry specifics of sub-portfolios*

| Basic segment | | Contracted segment | | The entire corporate portfolio | |
|---|---|---|---|---|---|
| $R^2$ **Stats** | | | | | |
| 85,5% | | 84,8% | | 84,8% | |
| **Breusch - Pagan Stats** (Breusch T. S., Pagan A. R., 1979) (norm -up to 10%) | | | | | |
| 7,53% | | 6,01% | | 8,85% | |
| **Variables** | **Coefficients** | **Variables** | **Coefficients** | **Variables** | **Coefficients** |
| Regression free term | 1,81 | Regression free term | 4,19 | Regression free term | 4,69 |
| GDP for the quarter, billion RUR, current prices | −0,000052 | The real salary of one worker | −4,84 | The real salary of one worker | −5,75 |
| USD LIBOR 1 year | 0,34 | RTS Index | −0,00073 | USD LIBOR 1 year | 0,36 |
| The real salary of one worker | −4,05 | Fixed capital investments,  with a lag of 1 year | −1,50 | MICEX Index, | −0,00051 |
| RTS Index | −0,00040 | | | The real salary of one worker with a lag of 1 year | −1,67 |

The macro model predicts the behavior of $PD_i$ in the future under a certain baseline scenario supplied by the bank's special analytical service or official sources (Bank of Russia, Ministry of Finance, World Bank, etc.). Table 2 shows that macro models have a fairly high determinism ratio of $R^2$ with low heteroscedasticity. This is the main advantage of building a macro model on a number of $PD_i$, built on macro-economic data, taking into account the industry specifics of the bank's portfolio. If you directly build a macro model on the local data of the bank $DF_i$ the determination ratio is worse.

## 6. Conclusion, Contribution and Implication

From the study presented in this paper, we can draw the following conclusions:

- The model of filtering the probability of default from the data of the Bank of Russia on delay was confirmed on the data of the NCB on defaults of individuals;

- The bank's data is updated monthly and is a reliable source including for audit;

- The probability of default and its volatility depends significantly on the industry;

- Portfolio segments for which a macro model is built are formed taking into account the industry shares of their own portfolio;

- The quality of regression macro models of market-oriented segments was quite high;

The DF (curve) of own credit portfolio may differ from the quasi-DF market, but you should expect a high correlation between them. There may be a significant idiosyncratic component associated with internal processes. For small and medium-sized banks presented a method of building a macro model, based on filtering the probability of default from the data of the Bank of Russia, but taking into account the industry specifics of the bank's loan portfolio and its own average frequency of realized defaults over a long period is probably the only audited method of building a macro model, not contrary to international standards IFRS 9[9] . Trying to build a macro model on your own data may not be valid because of the low statistical significance. For large banks with sufficient default statistics to build a statistically significant exogenous series of $DF_i$, building a regression macro model based on it may have low discriminatory ability due to the subjective factors affecting DF.

---

[9]International Financial Reporting Standard (IFRS) 9 "Financial Instruments" introduced as mandatory for Russian banks from 2018. https://www.minfin.ru/common/upload/library/2017/02/main/MSFO_IFRS_9_1.pdf

## 7. References

1. Kuznecov K.B., Malahova T.A., SHimanovskij K.V. (2011) Metody ocenki veroyatnosti defolta otraslej ekonomiki dlya celej bankovskogo nadzora// Vestnik permskogo universiteta. EKONOMIKA, Vol. 1(8), S.71-78 (In Russian)

2. Merton, R.C., (1974), "On the Pricing of Corporate Debt: The Risk Structure of Interest Rate", Journal of Finance 29, 449-470.

3. McQuown, J.A., (1993), "A Comment On Market vs. Accounting Based Measures of Default Risk", KMV Corporation.

4. Sobehart, J. R., Stein, R. M., Mikityanskaya, V., Li, L, (2000), "Moody's Public Firm Risk Model: A Hybrid Approach to Modeling Short-Term Default Risk", Moody's Investors Service (February).

5. Lapshin V. A., Smirnov S. N. (2012) Konsolidaciya i agregaciya ocenok veroyatnosti defolta// Upravlenie riskom. T. 61-63. № 1-3. S. 14-44. (In Russian)

6. Hodrick, Robert; Prescott, Edward C. (1997). "Postwar U.S. Business Cycles: An Empirical Investigation". Journal of Money, Credit, and Banking. 29 (1): 1–16. doi:10.2307/2953682

7. Breusch, T. S.; Pagan, A. R. (1979). "A Simple Test for Heteroskedasticity and Random Coefficient Variation". Econometrica. 47 (5): 1287–1294. doi:10.2307/1911963